\documentstyle[emulateapj,psfig,flushrt,tighten]{article}
\def\munu{_{\mu \nu}}
\def\kms{\,{\rm km\,s^{-1}}}

\def\msun{\,{\rm M_\odot}}

\def\etal{{et al.\ }}

\newcommand\beq{\begin{equation}}
\newcommand\eeq{\end{equation}}
\newcommand{\ba}{\begin{eqnarray}}
\newcommand{\ea}{\end{eqnarray}}
\def\spose#1{\hbox to 0pt{#1\hss}}
\def\lta{\mathrel{\spose{\lower 3pt\hbox{$\mathchar"218$}}
     \raise 2.0pt\hbox{$\mathchar"13C$}}}
\def\gta{\mathrel{\spose{\lower 3pt\hbox{$\mathchar"218$}}
     \raise 2.0pt\hbox{$\mathchar"13E$}}}
\def\ni{\noindent}

\righthead{Gravitational Waves from Massive Black Hole Binaries}
\submitted{ApJ, in press}
\makeatletter

\newenvironment{figurehere}
  {\def\@captype{figure}}
  {}
\makeatother
\begin{document}

\title{Low-frequency gravitational radiation from coalescing massive black hole binaries 
in hierarchical cosmologies}

\author{Alberto Sesana\altaffilmark{1}, Francesco Haardt\altaffilmark{1},
Piero Madau\altaffilmark{2}, \& Marta Volonteri\altaffilmark{2}}

\altaffiltext{1}{Dipartimento di Fisica \& Matematica, Universit\'a dell'Insubria, via 
Valleggio  11, 22100 Como, Italy.}
\altaffiltext{2}{Department of Astronomy \& Astrophysics, University of 
California, 1156 High Street, Santa Cruz, CA 95064.}

\begin{abstract}
We compute the expected low-frequency gravitational wave signal from coalescing 
massive black hole (MBH) binaries at the center of galaxies in a hierarchical 
structure formation scenario in which seed holes of intermediate mass form 
far up in the dark halo ``merger tree''. The merger history of dark matter 
halos and associated MBHs is followed via cosmological Monte Carlo 
realizations of the merger hierarchy from redshift $z=20$ to the present in a 
$\Lambda$CDM cosmology. MBHs get incorporated through halo mergers into 
larger and larger structures, sink to the center owing to dynamical friction
against the dark matter background, accrete cold material in the merger 
remnant, and form MBH binary systems. Stellar dynamical (three-body) 
interactions cause the hardening of the binary at large separations, while 
gravitational wave emission takes over at small radii and leads to the final 
coalescence of the pair. A simple scheme is applied in which the ``loss cone'' 
is constantly refilled and a constant stellar density core forms because of the 
ejection of stars by the shrinking binary. The integrated emission from 
inspiraling MBH binaries at all redshifts is computed in the quadrupole 
approximation, and results in a gravitational wave background (GWB) 
with a well defined shape that reflects the 
different mechanisms driving the late orbital evolution. The characteristic 
strain spectrum has the standard $h_c(f)\propto f^{-2/3}$ behavior only in 
the range $f=10^{-9}-10^{-6}\,$Hz. At lower frequencies the orbital decay 
of MBH binaries is driven by the ejection of background stars (``gravitational
slingshot''), and the strain amplitude increases with frequency, $h_c(f)\propto
f$. In this range the GWB is dominated by $10^9-10^{10}\,$M$_{\odot}$ 
MBH pairs coalescing at $0\lta z\lta 2$. At higher frequencies, $f>10^{-6}\,$Hz,
the strain amplitude, as steep as $h_c(f)\propto f^{-1.3}$, is 
shaped by the convolution of last stable circular orbit emission by 
lighter binaries ($10^{2}-10^{7}\,$M$_\odot$) populating galaxy halos at all 
redshifts. We discuss the observability of inspiraling MBH binaries by a 
low-frequency 
gravitational wave experiment such as the planned {\it Laser Interferometer 
Space Antenna} ({\it LISA}). Over a 3-year observing period {\it LISA} 
should resolve this GWB into discrete sources, detecting $\approx 60$ ($\approx 250$)
individual events above a $S/N=5$ ($S/N=1$) confidence level. 
\end{abstract}


\section{Introduction \label{intro}}
Studies of gravitational wave (GW) emission and of its detectability are 
becoming increasingly topical in astrophysics. Technological developements of 
bars and interferometers bring the promise of a future direct observation of 
gravitational radiation, allowing to test one of the most fascinating 
prediction of General Relativity, and, at the same time, providing a new 
powerful tool in the astronomical investigation of highly relativistic 
catastrophic events, such as the merging of compact binary systems and the 
collapse of massive stellar cores.

Massive black hole (MBH) binaries are among the primary candidate sources of 
GWs at mHz frequencies (see, e.g., Haehnelt 1994; Jaffe \& Backer 2003; Wyithe
\& Loeb 2003), the range probed by the space-based {\it Laser 
Interferometer Space Antenna} ({\it LISA}). Today, MBHs are ubiquitous in the 
nuclei of nearby galaxies (see, e.g., Magorrian et al. 1998). If MBHs were 
also common in the past (as implied by the notion that many distant galaxies 
harbor active nuclei for a short period of their life), and if their host 
galaxies experience multiple mergers during their lifetime, as dictated by 
popular cold dark matter (CDM) hierarchical cosmologies, then MBH binaries 
will inevitably form in large numbers during cosmic history. MBH pairs that 
are able to coalesce in less than a Hubble time will give origin to the 
loudest GW events in the universe.

Provided wide MBH binaries do not ``stall'', their GW driven inspiral will 
follow the merger of galaxies and pregalactic structures at high redshifts. 
A low-frequency detector like {\it LISA} will be sensitive to waves from coalescing binaries 
with total masses in the range $10^3-10^6\,\msun$ out to $z\sim 5-10$ (Hughes 
2002). Two obvious outstanding questions are then how far up in the dark halo
merger hierarchy do MBHs form, and whether stellar dynamical processes 
can efficiently drive wide MBH binaries to the GW emission stage.
The luminous $z\approx 6$ quasars discovered in the Sloan Digital Sky 
Survey (Fan \etal 2001) imply that black holes more massive than a few 
billion solar masses were already assembled when the universe was less 
than a billion years old. These MBHs could arise naturally from the growth by 
Eddington-limited gas accretion of seed stellar-mass or intermediate-mass 
holes forming at $z\gta 10$ (Haiman \& Loeb 2001). It seems also clear that, 
following the merger of two halo$+$MBH systems of comparable mass (``major 
mergers''), dynamical friction will drag in the ``satellite'' halo (and 
its MBH) toward the center of the 
more massive progenitor: this will lead to the formation of a bound MBH 
binary in the violently relaxed core of the newly merged stellar system.
What remains uncertain is the demography of the MBH population 
as a function of redshift and the dynamics of MBH binaries, i.e. whether
binaries will ``hung up'' before the backreaction from GW emission becomes
important.    

In this paper we study the expected gravitational wave signal
from inspiraling binaries in a hierarchical structure formation scenario 
in which seed holes of intermediate mass form far up in the dark halo 
``merger tree''. The model has been discussed in details in Volonteri, Haardt 
\& Madau (2003, hereafter Paper I). Seed holes are placed within rare 
high-density regions (minihalos) above the cosmological Jeans and cooling 
mass at redshift 20. Their evolution and growth is followed through Monte 
Carlo realizations of the halo merger hierarchy combined with semi-analytical
descriptions of the main dynamical processes, such as dynamical friction
against the dark matter background, the shrinking of MBH binaries via 
three-body interactions, their coalescence due to the emission of 
gravitational waves, and the ``gravitational rocket''. Major mergers lead to 
MBH fueling. The long dynamical 
frictional timescales leave many MBHs ``wandering'' in galaxy halos after a 
minor merger. Bound pairs form after major mergers and lose orbital angular 
momentum by capturing stars passing 
within a distance of the order of the binary semi-major axis and ejecting
them at much higher velocities. The 
heating of the surrounding stars by a decaying MBH pair creates a low-density 
core out of a preexisting stellar cusp, slowing down further binary hardening 
(see, e.g., Milosavljevic \& Merritt 2001). The model reproduces 
rather well the observed luminosity function of optically-selected quasars in 
the redshift range $1<z<5$ (Paper I), and provides a quantitative explanation 
to the stellar cores observed today in bright ellipticals as a result of the 
cumulative eroding action of shrinking MBH binaries (Volonteri, Madau, 
\& Haardt 2003, hereafter Paper II). 

The paper is organized as follows. In \S~2 we describe our merger tree 
algorithm coupled with semi-analytical recipes designed to track the dynamical 
history of MBHs in a cosmological framework. We compute black hole binary 
coalescence rates as a function of redshift and mass, and show that MBH 
binaries will shrink to the gravitational wave emission stage by scattering 
off background stars from stellar cusps. In \S~3 we summarize the basic 
theory of GW emission. In \S~4 we compute the cosmological GW background 
(GWB) from inspiraling MBH binaries and discuss the observability of 
individual coalescence events by {\it LISA}. Finally, we summarize our main 
results in \S~5. Unless otherwise stated, all results shown below refer to 
the currently favoured $\Lambda$CDM world model with $\Omega_M=0.3$, 
$\Omega_\Lambda=0.7$, $h=0.7$, $\Omega_b=0.045$, $\sigma_8=0.93$, and $n=1$.

\section{MBH binaries\label{trees}}

\subsection{Assembly and growth of MBH\lowercase{s}}

Following Papers I and II, we track backwards the merger history of parent
halos with present-day masses in the range $10^{11}-10^{15}\,\msun$ 
with a Monte Carlo algorithm based on the extended Press-Schechter formalism
(see, e.g., Cole \etal 2000). The tree is averaged over a large number of  
different realizations, for a total of 220 simulated halos, 
and the results weighted by the Press-Schechter halo
mass function at the chosen output redshift. Pregalactic seed holes form 
with intermediate 
masses ($m_\bullet=150\,\msun$) as remnants of the first generation of massive 
metal-free stars with $m_*>260\,\msun$ that do not disappear as 
pair-instability supernovae (Madau \& Rees 2001). We place them in isolation 
within halos above $M_{\rm seed}=1.6\times 10^7\,\msun$ collapsing 
at $z=20$ (the highest redshift computational costs allow us to follow the 
merger hierarchy to) from rare $>3.5\sigma$ peaks of the primordial density 
field. While $Z=0$ stars with $40<m_*<140\,\msun$ are also predicted 
to collapse to MBHs with masses exceeding half of the initial stellar mass 
(Heger \& Woosley 2002), we find that the GWB signal in the {\it LISA}
window is not very sensitive to the precise choice for the seed hole mass.

Hydrodynamic simulations of major mergers have shown that a significant 
fraction of the gas in interacting galaxies falls to the center of the merged 
system (Mihos \& Hernquist 1996): the cold gas may be eventually driven 
into the very inner regions, fueling an accretion episode and the growth of 
the nuclear MBH (see, e.g. Kauffmann \& Haehnelt 2000). Since the local MBH 
mass density is consistent with the integrated luminosity density of quasars 
(Yu \& Tremaine 2002), the fraction of cold gas ending up in the hole must 
depend on the properties of the host halo in such a way to ultimately lead 
to the observed correlation between stellar velocity dispersion and SMBH mass.
As in Papers I and II, we assume that in each major 
merger (defined here as a merger between halos of mass ratio $>0.3$),
the more massive hole accretes, at the Eddington rate, a gas mass that scales 
with the fifth power of the circular velocity of the host halo,
\begin{equation}
\Delta m_{\rm acc}=3.6\times 10^6\,\msun~{\cal K}\,V_{c,150}^{5.2},
\label{macc_eq}
\end{equation}
where $V_{c,150}$ is the circular velocity of the merged system in units of
150 $\kms$. 
This relation follows from combining the $m_{\rm BH}-\sigma_*$ and
the $V_c-\sigma_*$ relations given in Ferrarese (2002, see Papers I and II). 
The normalization factor $\cal K$ is of order unity and is fixed
a posteriori in order to reproduce the correlation between stellar velocity 
dispersion and nuclear black hole mass observed today in nearby galaxies 
(Ferrarese \& Merritt 2000; Gebhardt \etal 2000).

Inside a halo of mass $M$, the dark matter is distributed according to a NFW 
profile (Navarro, Frenk, \& White 1997),
\begin{equation}
\rho_{\rm DM}(r)={M\over 4 \pi r (r+r_{\rm vir}/c)^2 f(c)},
\label{eqn:nfw1}
\end{equation}
where $r_{\rm vir}$ is the virial radius, $c$ is the halo concentration 
parameter and $f(c)=\ln(1+c)-c/(1+c)$. The distribution of concentrations 
and its scaling with halo mass and redshift of collapse are taken from 
Bullock \etal (2001). Dynamical friction drags satellite progenitors -- and 
their MBHs if they host one or more -- towards the center of the more massive 
preexisiting system, on a timescale that depends on the orbital parameters 
of the satellites (van den Bosch \etal 1999) and on tidal mass loss/evaporation
(Taffoni \etal 2003). The initial occupation fraction of seed holes is so 
small that the merging of two halos both hosting a black hole is a rare event
even with the assumed ``bias'', and MBHs evolve largely in isolation. At 
redshifts below 15, however, more than 10\% of hosts contain two or more 
MBHs, and bound binaries start forming in significant numbers. 

The evolution of a bound binary is determined by the initial central stellar 
distribution. We model this as a (truncated) singular isothermal sphere (SIS) 
with one-dimensional velocity dispersion $\sigma_*$ and density 
\beq 
\rho_*(r)={\sigma_*^2\over 2\pi G r^2}.
\eeq
The stellar velocity dispersion is related to the halo circular velocity 
$V_c$ at the virial radius following Ferrarese (2002), 
\beq
\log V_c=(0.88\pm 0.17)\log \sigma_*+(0.47\pm 0.35).
\label{vsigma}
\eeq

\subsection{Binary merger rates}

The semimajor axis $a(t)$ of a bound binary continues to shrink owing to 
dynamical friction from distant stars acting on each MBH individually, until 
it becomes ``hard'' when the separation falls below (Quinlan 1996)
\beq
a_h={Gm_2\over 4\sigma_*^2}=0.2\,{\rm pc}~m_6\,\sigma_{70}^{-2},
\eeq
where $m_2(<m_1)$ is the mass of the lighter hole, $m_6\equiv m_2/10^6\,
\msun$, and $\sigma_{70}$ is the stellar velocity dispersion in units of 
$70\,\kms$. 
After this stage the MBH pair hardens via three-body interactions, i.e., by 
capturing and ejecting at much higher velocities the stars passing by within 
a distance $\sim a$ (Begelman, Blandford, \& Rees 1980). We assume that the 
``bottleneck'' stages 
of binary shrinking occur for separations $a<a_h$; during a galactic merger, 
after a dynamical friction timescale, we place the MBH pair at $a_h$ and let 
it evolve. The hardening of the binary modifies the stellar density profile,
removing mass interior to the binary orbit, 
depleting the galaxy core of stars, and slowing down further hardening. 
If ${\cal M}_{\rm ej}$ is the stellar mass ejected by the pair, 
the binary evolution and its effect on the galaxy core are determined by 
two dimensionless quantities: the hardening rate
\beq
H={\sigma_*\over G\rho_*}{d\over dt}{1\over a},
\label{eqH}
\eeq
and the mass ejection rate
\beq
J={1\over (m_1+m_2)}\,{d{\cal M}_{\rm ej}\over d\ln(1/a)}.
\label{eqJ}
\eeq
The quantities $H$ and $J$ can be found from scattering experiments
that treat the test particle-binary encounters one at a time (Quinlan 1996).
We assume as in Papers I and II (see also Merritt 2000) that 
the stellar mass removal creates a core of radius $r_c$ and constant 
density $\rho_c\equiv \rho_*(r_c)$, so that the total mass ejected as
the binary shrinks from $a_h$ to $a$ can be written as
\begin{equation}
{\cal M}_{\rm ej}= {2\sigma_*2\over G}(r_c-r_i)+M_i-M_c
= {4\over 3}{\sigma_*^2 (r_c-r_i)\over G}, 
\label{rcore}
\end{equation}
where 
$M_c=4\pi\rho_cr_c^3/3$, $M_i=4\pi\rho_ir_i^3/3$, 
$r_i=r_c(t=0)$ is the radius of the (preexisting) core when 
the hardening phase starts at $t=0$, and $\rho_i\equiv \rho_*(r_i)$. 
The core radius then grows as 
\begin{equation}
r_c(t)=r_i+{3\over4 \sigma_*^2}G(m_1+m_2)\int_{a(t)}^{a_h}{\frac{J(a)}{a}\,da}.
\label{rc}
\end{equation}
The binary separation quickly falls below $r_c$ and subsequent evolution 
is slowed down due to the declining stellar density, with a hardening time,
\beq
t_h=|a/\dot a|={2\pi r_c(t)^2\over H\sigma_*a}, \label{thard}
\eeq
that becomes increasingly long as the binary shrinks. The mass ejected
increases approximately logarithmically with time, and the binary ``heats'' 
background stars at radii $r_c\gg a$. We assume that the effect
of the hierarchy of MBH binary interactions is cumulative (the ``core 
preservation'' model of Paper II), i.e. the mass displaced by the pair is 
not replenished after
every major merger event and $r_i$ continues to grow during 
the cosmic evolution of the host. This is supported by N-body simulations
involving mergers of spherical galaxy models with different density
profiles, and
showing that the remnant profile is quite close to the profile of the
progenitors -- in other words that the core appears to be preserved
during such mergers (e.g. Fulton \& Barnes 2001).
The above scheme can be modified by rare triple black hole interactions, when
another major merger takes place before the pre-existing binary has had time 
to coalesce (see, e.g., Hut \& Rees 1992). In this case
there is a net energy exchange between the binary and the third incoming black 
hole, resulting in the ejection of the lighter hole and the recoil of the 
binary. The binary also becomes more tightly bound. The reader is referred 
to Paper I for a detailed discussion of the role played by triple interactions. 
If hardening continues down to a separation
\beq
a_{\rm gw}= 0.0014\,{\rm pc}~~\left[{(m_1+m_2)m_1m_2\over 10^{18.3}\,\msun^3
}\right]^{1/4}\,t_9^{1/4},
\eeq
the binary will coalesce within $t_9$ Gyr due to the emission of 
gravitational waves. An equal mass pair must then manage to shrink by a factor 
\beq
{a_h\over a_{\rm gw}}\approx 150~m_6^{3/4}\,\sigma_{70}^{-2}\,t_9^{-1/4}
\eeq 
for gravity wave emission to become efficient. Since the hardening and mass 
ejection rate coefficients are $H\approx 15$ and $J\approx 1$ in the limit of 
a very hard binary (Quinlan 1996), the hardening time in equation 
(\ref{thard}) can be rewritten (for $m_1=m_2$ and $r_i\rightarrow 0$) as
$$
t_h\approx 4.7\times 10^4\,{\rm yr}~m_6\,\sigma_{70}^{-3}\,(a_h/a)\,
\ln^2(a_h/a)
$$
\beq
~~~~~\approx 5.6\times 10^4\,{\rm yr}~m_6^{0.35}\,(a_h/a)\,\ln^2(a_h/a),
\label{lnth}
\eeq
where the second equality assumes the $m_{\rm BH}-\sigma_*$ relation 
of Ferrarese (2002). Allowing for the cumulative effect of a hierarchy of 
MBH binary interactions (i.e. for the existence of a pre-existing core
of radius $r_i$) increases somewhat the hardening time in equation 
(\ref{lnth}). Yet, we find $t_h$ to be comparable or shorter than the then 
Hubble time at all but the highest redshifts. Figure \ref{fig1} shows the 
number of MBH binary coalescences per unit redshift per unit {\it observed} year,
$dN/dzdt$, predicted by our model. The observed event rate is obtained by 
dividing the rate per unit proper time by the $(1+z)$ cosmological time 
dilation factor. Each panel shows the rate for 
different $m_{\rm BH}=m_1+m_2$ mass intervals, and lists the integrated 
event rate, $dN/dt$, across the entire sky. The number of events per observed
year per unit redshift peaks at $z=2$ for $10^7<m_{\rm BH}<10^9\,\msun$, at 
$z=3-4$ for $10^5<m_{\rm BH}<10^7\,\msun$, and at $z=10$ for 
$m_{\rm BH}<10^5\,\msun$, i.e. the lower the black hole mass, the higher the
peak redshift. Beyond the peak, the event rate decreases steeply
with cosmic time. Previous calculations of this quantity have either
assumed instantaneous black hole coalescence and computed halo merger rates only
out to $z\sim 5$ (Menou, Haiman, \& Narayanan 2001), or have assigned a 
constant (independent of black hole mass, background stellar density, and 
redshift) efficiency factor to the coalescence process (Wyithe \& Loeb 2003),
or have used phenomenological and empirical rates (Jaffe \& Backer 2003).      
Our rates are significantly smaller than those computed by Wyithe \& Loeb 
(2003), consistent with their assumption that all MBH binaries actually 
coalesce.   
\begin{figurehere}
\vspace{0.5cm}
\centerline{\psfig{figure=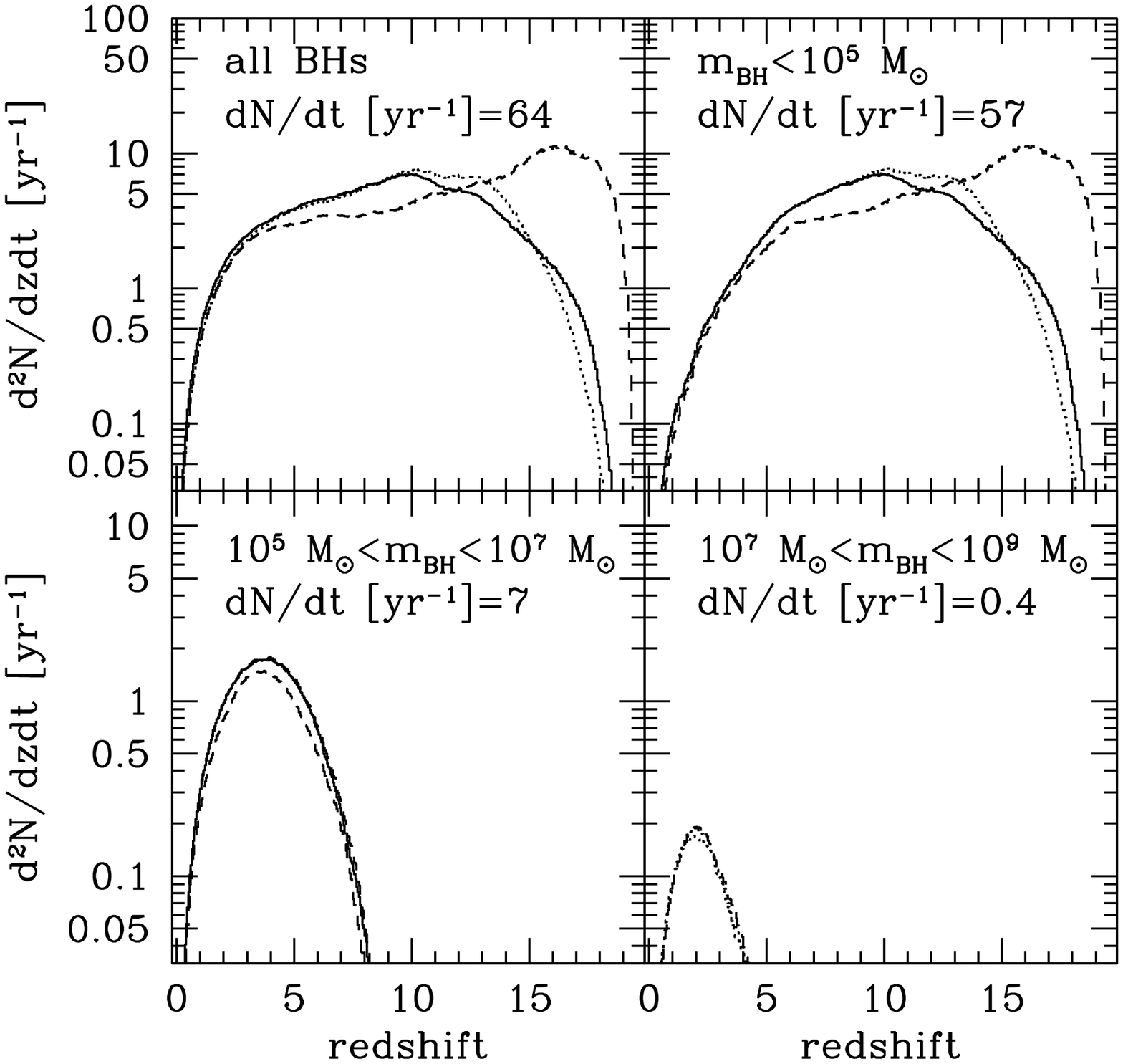,width=3.6in}}
\vspace{-0.0cm}
\caption{\footnotesize Number of MBH binary
coalescences observed per year at $z=0$, per unit redshift, in different
$m_{\rm BH}=m_1+m_2$ mass intervals. Each panel also lists the
integrated event rate, $dN/dt$, predicted by our model. The rates ({\it
solid lines}) are
compared to a case in which triple black hole interactions are switched off
({\it dotted lines}). Triple hole interactions increase the coalescence rate
at very high redshifts, while, for $10<z<15$, the rate is decreased because
of the reduced number of surviving binaries. {\it Dashed lines}: rates
computed assuming binary hardening is instantaneous, i.e. MBHs coalesce
after a dynamical friction timescale.
}
\label{fig1}
\vspace{+0.5cm}
\end{figurehere}

While our calculations are the first to include a detailed model of MBH 
binaries dynamics, we note that our scheme implicitly neglects the depopulation 
of the ``loss cone''(Frank \& Rees 1976), since it is the total stellar density 
that is allowed to decrease following equation (\ref{rc}), not the density of 
low angular momentum stars. The effect of loss-cone depletion (the depletion of 
low-angular momentum stars that get close enough to extract energy from a 
hard binary) has traditionally been one of the major uncertainties in 
constructing viable merger scenarios for MBH binaries (Begelman \etal 1980). 
Several processes have been recently studied that should all mitigate
the problems associated with loss-cone depletion, such as the wandering of the 
binary center of mass from the galaxy center induced by continuous interactions 
with background stars (Chatterjee, Hernquist, \& Loeb 2003), the large supply 
of low-angular momentum stars in significantly flattened or triaxial galaxies 
(Yu 2002), the presence of a third hole (Blaes, Lee, \& Socrates 2002), and
the randomization of stellar orbits due to the infall of small satellites 
(Zhao, Haehnelt, \& Rees 2001). Our scheme also neglects the role
of gaseous -- rather than stellar dynamical -- processes in driving the 
evolution of a MBH binary (see, e.g., Escala \etal 2004). We will try to 
address the effect of some these uncertainties on the predicted GWB in \S~4.

There is another process that our hierarchical MBH evolution scenario includes: 
this is the ``gravitational rocket'', the recoil due to the non-zero 
net linear momentum carried away by GWs in the coalescence of two unequal mass 
black holes. Radiation recoil is a strong field effect that depends on the lack 
of symmetry in the system, and may diplace MBHs from galaxy centers or even 
eject them into intergalactic space (Redmount \& Rees 1989). Its outcome is 
still uncertain, as fully general relativistic numerical computations of 
radiation reaction effects are not available at the moment. Here we have 
used the extrapolated perturbative results on test mass particles of Fitchett 
\& Detweiler (1984), predicting in a Schwarzschild geometry recoil velocities 
larger than $200\,\kms$ only for mass ratios $m_2/m_1>0.4$. As shown in 
Figure \ref{fig2}, in our model the typical mass ratio of merging binaries 
exceeds 0.4 at $z>10$. It is at early times then that the gravitational rocket 
is effective at ejecting MBHs from the shallow potential wells (escape 
velocities less than $100\,\kms$) of their hosts (Madau \etal 2004). 
Overall, we find that $\approx 25\%$ of coalesced pairs recoil and escape their host
halos at $z \sim 10$, and that the ejected holes have typical masses of $\gta 1000\, \msun$. 
The detectability of recoiling MBHs has been recently addressed by Madau \& Quataert (2004).

\section{Gravitational wave signal \label{GW}}

\subsection{Basic theory \label{GWbasic}}

The theory of GW emission in the quadrupole approximation is textbook 
topic (e.g., Schutz \& Ricci 2001; Maggiore 2000), and we recall
here only the basic relations. The Einstein equations can be linearized over 
the perturbed Minkowski metric. The perturbed part of the metric or ``strain''
admits a wave solution out of the source of the stress-energy tensor. Such 
solution becomes particulary easy to treat in the so-called transverse-traceless
(TT) gauge. In the TT gauge the strain at distance $r$ from the source has 
the form
\beq
\begin{array} {l}
h_{\mu 0}^{TT}=0 \\
h_{ik}^{TT} (t,r)={2G\over c^4 r}\left[
{d^2\over{dt^2}}Q_{ik}^{TT}\left( t-{r\over c} \right) \right], 
\end{array}
\label{strain}
\eeq
so that only the spatial components are not vanishing.
The term $Q_{ik}^{TT}$ is the reduced quadrupole momentum projected into the 
TT gauge, containing the stress-energy tensor of the gravitational source, 
and evaluated at the retarded time $t-r/c$.

\begin{figurehere}
\vspace{+0.5cm}
\centerline{\psfig{file=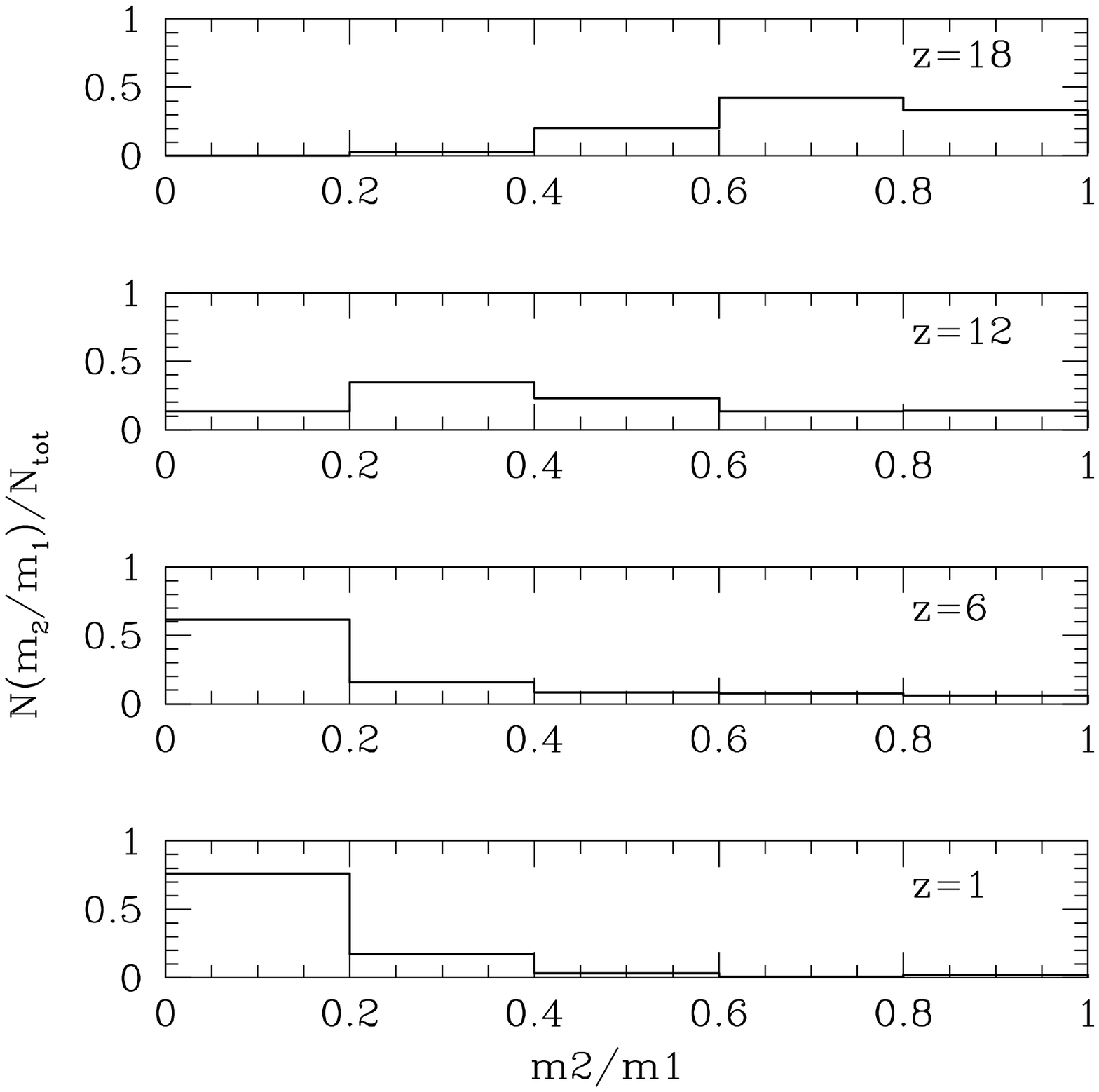,width=3.6in}}
\vspace{-0.0cm}
\caption{\footnotesize Normalized distribution of mass ratios of coalescing MBH binaries
at different epochs, averaged over 120 realizations. At very high redshift
binary members are seed holes with nearly equal masses; as time goes on MBHs
grow due to gas accretion, and low mass ratios become more probable.}
\label{fig2}
\vspace{+0.5cm}
\end{figurehere}

It is then possible to associate an energy momentum pseudo-tensor to the wave, 
which, to first perturbative order, transforms as a tensor. 
Such pseudo-tensor is
\beq
t\munu={c^4\over{32\pi G}}\left<h^{TT}_{ij,\mu}h^{TTij}_{   ,\nu}\right>
\label{pseudotensor}
\eeq
where $<>$ is an average over wavelength. If we consider a GW propagating 
along the positive $x$ direction, then the associated energy flux will be
\beq
ct_{01}={dE_{\rm gw}\over{dtdS}} ={c^5\over{32\pi G}}\left<h^{TT}_{ij,0}
h^{TTij}_{,1}\right>.
\label{fluxGW}
\eeq
Taking equation (\ref{strain}), deprojecting the quadrupole momentum 
$Q^{TT}$ out of the TT gauge to get $Q$, and integrating over the surface 
$S$, one derives the luminosity associated with GW emission:
\beq
{dE_{\rm gw}\over dt}=\int_S {dE_{\rm gw}\over{dtdS}}dS= {G\over{5c^5}}\left<{d^3\over{dt^3}}
Q_{ij}\,{d^3\over{dt^3}}Q^{ij}\right>
\label{lumGW}.
\eeq
This depends on the time third derivative of the (reduced) quadrupole 
momentum of the source.

\subsection{Gravitational waves from binary systems \label{GWbinary}}

In the stationary case, i.e., assuming no orbital decay, the GW emission 
spectrum of a MBH pair in a circular orbit of radius $a$ is a delta function 
at rest-frame frequency $f_r=\omega/\pi$, where $\omega=\sqrt{G(m_1+m_2)/a^3}$ 
is the Keplerian angular 
frequency of the binary. Orbital decay (due to GW emission and/or the ejection
of background stars) results in a shift of the emitted frequency to increasingly
larger values as the binary evolution proceeds. The energy spectrum integrated 
over the entire radiating lifetime of the source (a continuum formed by the 
superposition of delta functions) can be computed once the energy emitted
per logarithmic frequency interval,
\begin{equation}
{dE_{\rm gw}\over{d\ln{f_r}}}={dE_{\rm gw}\over{dt}}{dt\over{da}}
{da\over{df_r}}f_r,
\label{dEdf}
\end{equation}
is known. This is related to the characteristic strain, $h_c$, through
\begin{equation}
{dE_{\rm gw}\over{d\ln{f_r}}} \propto f_r^2 \, h^2_c(f_r),
\label{eqhc}
\end{equation}
where $h_c$ is defined as 
\begin{equation}
\langle h_{\alpha \beta}(t)\,h^{\alpha \beta}(t) \rangle = 2 \int (d\ln{f_r})\,
h^2_c(f_r).
\end{equation}
The characteristic strain is related to the Fourier transform of the strain 
$\tilde h(f_r)$, $h_c^2(f_r)=2f_r|\tilde h(f_r) |^2$. 
In the case of a binary system equation (\ref{lumGW}) can be rewritten as:
\begin{equation}
{dE_{\rm gw}\over dt}=\frac{32\pi^{10/3}\,G^{7/3}}{5c^5}({\mathcal M}f_r)^{10/3}.
\label{lumGWbin}
\end{equation}
The radiated power is an increasing function of frequency and of 
the ``chirp mass'' of the system, 
\begin{equation}
{\mathcal M} \equiv {{m_1^{3/5}m_2^{3/5}}\over{\left(m_1+m_2\right)^{1/5}}}.
\label{chirp}
\end{equation}
The term $da/df$ in equation (\ref{dEdf}) is readily obtained by differentiating 
the relation $f_r=\omega/\pi$,
\begin{equation}
{da\over{df_r}}=-{2\over3}\left[{G (m_1+m_2)\over \pi^2}\right]^{1/3} f_r^{-5/3}.
\label{dadf}
\end{equation}
Finally, the term $da/dt$ represents the orbital decay due to angular momentum
losses. When the backreaction from GW emission dominates, the orbit shrinks
at a rate
\begin{equation}
\left({{da}\over{dt}}\right)_{\rm gw}=-{64\over 5}{G^3\over {c^5 a^3}}
m_1m_2(m_1+m_2), 
\label{drdtGW}
\end{equation}
and the resulting spectrum of GWs is \begin{equation}
\left({dE_{\rm gw}\over{d\ln{f_r}}} \right)_{\rm gw}= {(\pi G)^{2/3}\over 3}\,
{\mathcal M}^{5/3}f_r^{2/3}.
\label{spectrumGW}
\end{equation}
In the early stages of binary evolution, however, angular momentum losses are 
driven by stellar dynamical processes, and the emerging GW spectrum will show 
a different scaling with frequency. The model outlined in the previous section 
leads to the following expression for the orbital decay rate in the 
``gravitational slingshot'' regime (eq. \ref{thard}):
\beq
\left({{da}\over{dt}}\right)_{\rm gs}=-{H\sigma_* a^2\over 2\pi r_c^2}.
\label{drdtSL}
\eeq
Assuming an initially cuspy profile ($r_i=0$) and in the limit of a very hard
binary ($J\approx 1$), the core radius (eq. \ref{rc}) created by the shrinking 
pair can be written as 
\beq
r_c(t)\approx {3\over 4\sigma_*^2}G(m_1+m_2)\,\ln(a_h/a), 
\eeq
and the hardening rate becomes proportional (neglecting the logarithmic 
dependence) to
\begin{equation}
\left({{da}\over{dt}}\right)_{\rm gs}\propto \sigma_*^5 (m_1+m_2)^{-2}\,a^2.
\end{equation}
Contrary to the case when GW losses dominates, the binary decay slows down with
time because of the declining stellar density, and the rate $|da/dt|$ decreases 
as the orbit shrinks.
Inserting the above expression into equation (\ref{dEdf}) yields the following 
scaling for the emitted GW spectrum in the stellar ejection regime:
\begin{equation}
\left({dE_{\rm gw}\over{d\ln{f_r}}}\right)_{\rm gs}\propto m_1^2m_2^2(m_1+m_2)\,
\sigma_*^{-5}\,f_r^4.
\label{spectrumSL}
\end{equation}
The GW spectrum arising from a binary system during the slow {\it inspiral} 
phase spans a broad but finite interval of emitted frequencies, $f_{\rm min}<
f_r<f_{\rm max}$. We assume that the binary starts emitting GWs at separation 
$a=a_h$, so that the lower limit $f_{\rm min}$ is
\begin{equation}
f_{\rm min}=\frac{8\sigma_*^3}{\pi G} (m_1+m_2)^{1/2}\, m_{2}^{-3/2}.
\label{fmin}
\end{equation}
The upper limit $f_{\rm max}$ is set by the frequency emitted at $a=6Gm_1/c^2$, 
the radius of the closest stable circular orbit for two non-rotating holes,
\begin{equation}
f_{\rm max}={{c^3}\over{6^{3/2} \pi G}}\, (m_1+m_2)^{1/2}\, m_{1}^{-3/2}.
\label{fmax}
\end{equation}
We do not address in this work the higher frequency radiation emitted during 
the subsequent {\it coalescence} and {\it ringdown} phases (Flanagan \& Hughes 
1998). Figure \ref{fig3} shows three 
examples of GW spectra emitted by inspiraling binaries. The two regimes, 
stellar slingshot at low frequencies and GW losses at high frequencies, are 
clearly recognizable.

\begin{figurehere}
\vspace{0.5cm} \centerline{\psfig{figure=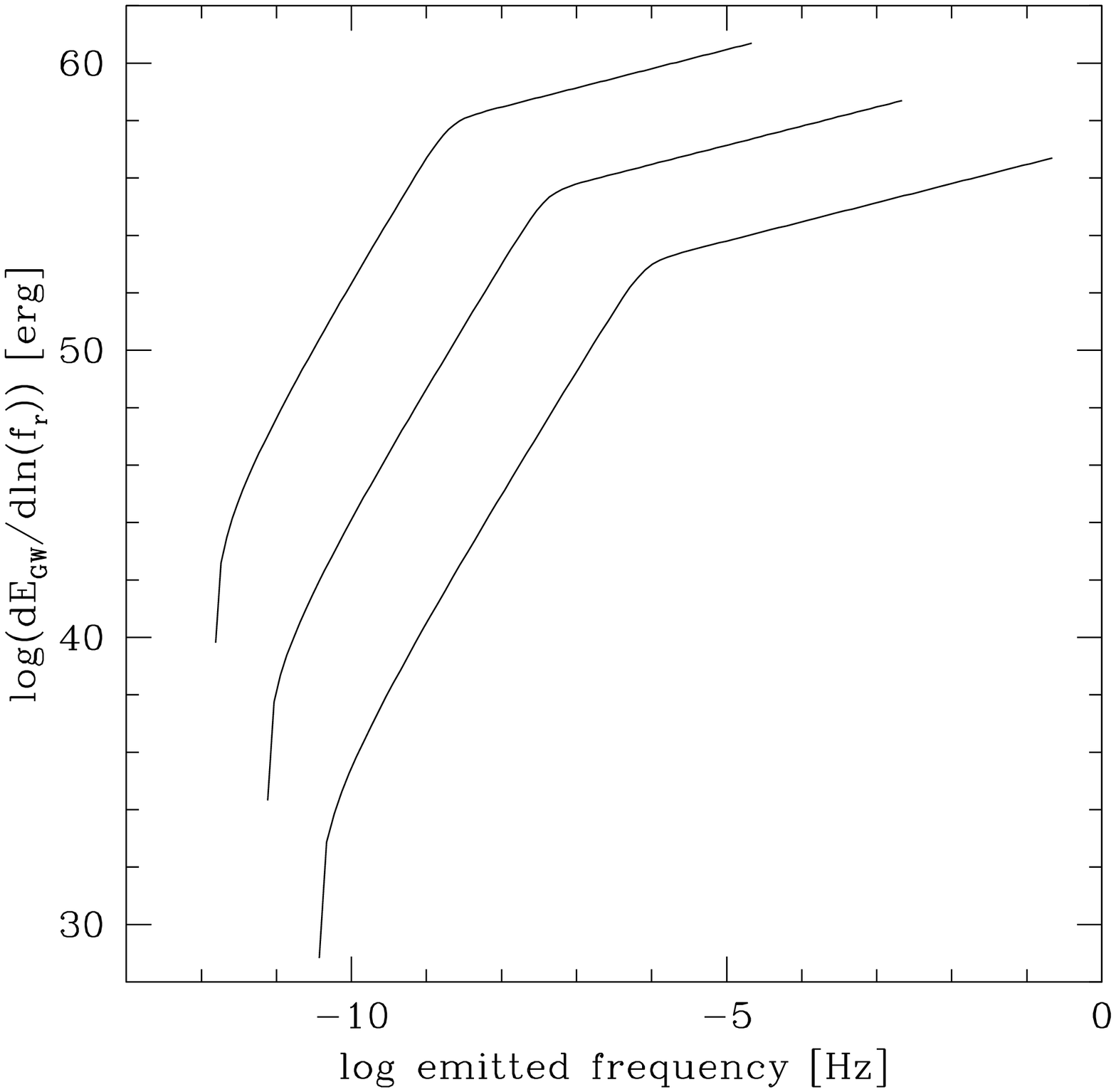,width=3.6in}}
\vspace{-0.0cm}
\caption{\footnotesize Energy per natural log frequency interval around $f_r$ emitted
in GWs by an equal mass ($m_1=m_2$) black hole binary system in a
stellar background of velocity dispersion $\sigma_*$. From top to bottom, the
three curves corresponds to $(m_1, \sigma_*)=(10^8\,\msun, 180\,\kms), (10^6\,
\msun, 66\, \kms)$, and $(10^4\,\msun, 24\,\kms)$, respectively.}
\label{fig3}
\vspace{+0.5cm}
\end{figurehere}

\subsection{Total background from cosmic sources \label{GWback}}

We can now compute the total background -- random GWs arising from a 
large number of independent events  -- produced by a population 
of sources at cosmological distances. Following 
Phinney (2001), the present-day energy density in GWs of frequency $f$ 
per natural log frequency interval is
\begin{equation}
c^2\rho_{\rm gw}(f)= {\pi \over 4}{c^2\over G} f^2h_c^2(f) =\int_0^{\infty} 
{dz\, {{N(z)}\over{1+z}}~{{dE_{\rm gw}} \over {d\ln{f_r}}}},
\label{pinnei}
\end{equation}
where $N(z)dz$ is the comoving number density of catastrophic events in the 
redshift interval $z$, $z+dz$, the factor $1/(1+z)$ accounts for the 
redshifting of gravitons, and the {\it observed} frequency $f$ is related
to the rest-frame frequency $f_r$ by  $f=f_r/(1+z)$. Equation 
(\ref{pinnei}) assumes that all sources emit the same intrinsic spectrum (i.e. 
all binaries have the same mass and shrink in the same stellar background),
and that the inspiraling timescale is much shorter than the then Hubble time,
i.e. the inspiral occurs at the same redshift as the coalescence. When 
binary decay occurs on timescales that are comparable or longer than the 
expansion time, equation (\ref{pinnei}) can be generalized as
\begin{equation}
c^2\rho_{\rm gw}(f)=f \int_0^{\infty} dz\, N(f_r,t)~{dE_{\rm gw}
\over dt}\,|{dt\over dz}|,
\label{pinneibis}
\end{equation}
where $N(f_r,t)$ is the comoving number density of binaries at cosmic time 
$t$ that radiate at frequencies between $f_r$ and $f_r+df_r$, 
$dt/dz=-(1+z)^{-1}H(z)^{-1}$ where $H(z)$ is the Hubble parameter, and 
$dE_{\rm gw}/dt$ is the GW luminosity at $f_r=f(1+z)$ 
measured in the rest-frame of the pair. The density of sources $N(f_r)$
is related to the space density of binaries with separations between $a$ and
$a+da$ by $N(f_r)df_r=-N(a)da$. If $\dot N_b$ is the birth rate (after 
dynamical friction) of MBH binaries, and $\dot N_c$ is the rate of 
coalescences, then $N(a)$ is given by the continuity equation
\beq
{\partial N\over \partial t}+{\partial (\dot a N)\over\partial a}=
\dot N_b\delta(a-a_h)-\dot N_c\delta(a).
\eeq
Note that, since in the situation we are 
considering the sources of GWs are MBH binaries with a mass function that 
changes with cosmic epoch, the term $N(f_r,t)\,dE_{\rm gw}/dt$ must be 
summed up over all binary types.  

\subsection{Resolvable signal \label{GWres}}

While the data stream of {\it LISA} will include confusion noise arising 
from a large number of unresolved galactic and extragalactic white dwarf 
binaries (Farmer \& Phinney 2003; Nelemans, Yungelson, \& Portegies-Zwart
2001), the GWB from MBH binaries will be resolvable into discrete 
sources at least in the high frequency end of the {\it LISA} sensitivity range.
The nature (stochastic or resolved) of the GWB can be assessed by counting
the number of events per resolution frequency bin whose observed signal 
is larger than a given sensitivity threshold $h_{c,{\rm min}}$. The background
is resolved if such number is less than unity and the cumulative signal 
from all sources below threshold is negligible. 

The resolution frequency bin is set by the minimum frequency resolvable by 
a given mission, and it is simply the inverse of the mission lifetime. 
The characteristic strain $h_c$ of a source at comoving coordinate distance 
$r(z)$ is 
\beq
h_c \simeq h \sqrt{n}, 
\eeq
where the strain amplitude $h$ (sky-and-polarization averaged) is given by
\beq
h={8\pi^{2/3}\over 10^{1/2}}{G^{5/3}{\cal M}^{5/3}\over c^4 r(z)}f_r^{2/3}, 
\label{eqthorne}
\eeq
and $n$ is the number of cycles a source spends at 
frequency $f_r$, i.e., $n=f_r^2/\dot f_r$ (Thorne 1987).
\footnote{The formula for $h$ in Thorne (1987) 
contains a multiplicative extra factor $\sqrt{4/3}$ to account for the Earth rotation.} 
Note that, for a finite observation time $\tau$, the number of cycles actually 
observed at frequency $f_r$ can not exceed $f\tau$. The frequency evolution 
$\dot f_r$ can be obtained by combining equations  
(\ref{dadf}) and (\ref{drdtGW}) in the GW regime, (\ref{dadf}) and 
(\ref{drdtSL}) in the slingshot regime. At the endpoint of the inspiraling phase, 
the number of cycles spent near that frequency is close to unity, so that $h_c \simeq h$.

\begin{figurehere}
\vspace{0.5cm} \centerline{
\psfig{figure=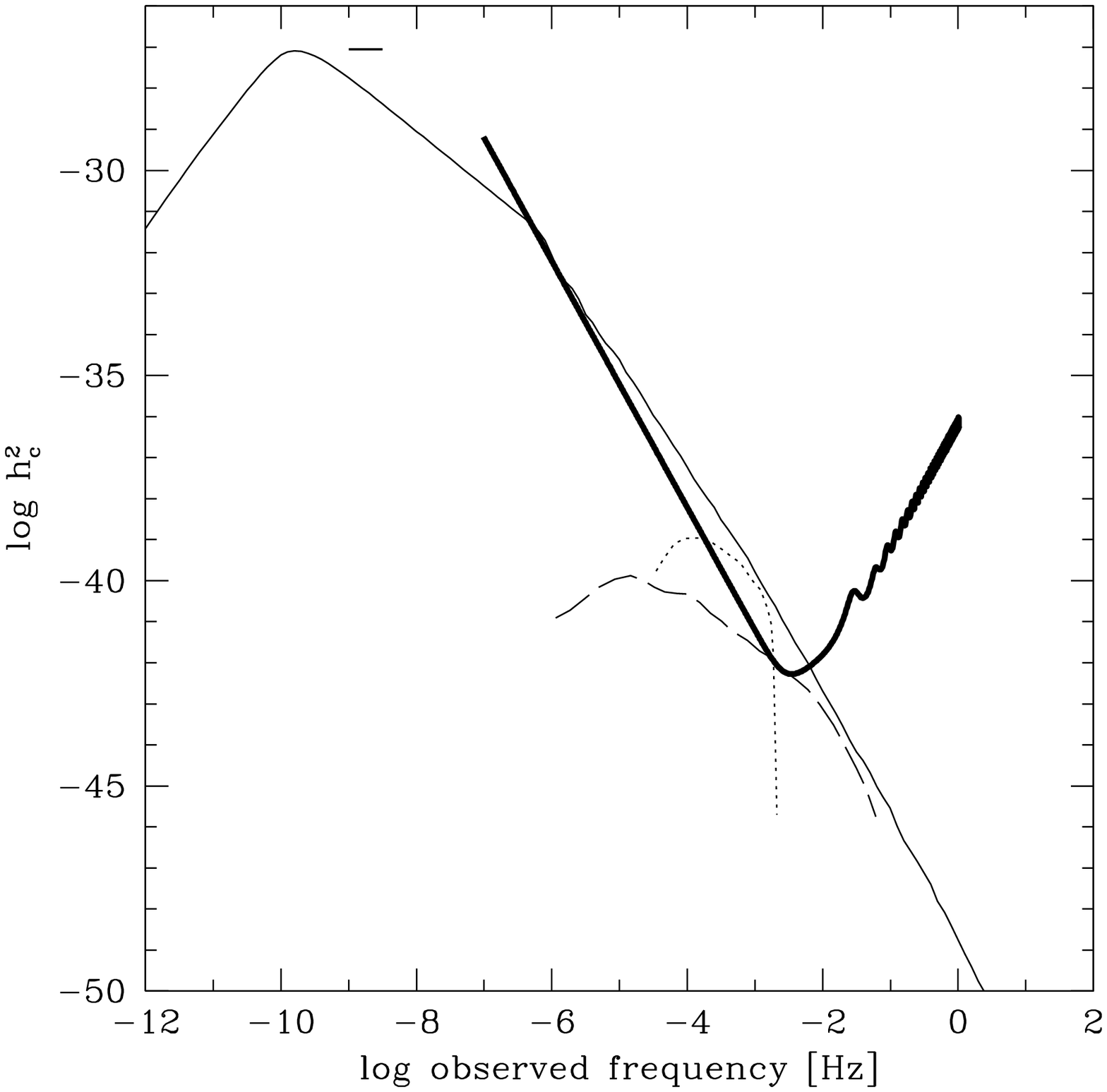,width=3.6in}}
\vspace{-0.0cm}
\caption{\footnotesize Gravitational wave background from inspiraling
MBH binaries in the hierarchical scenario described in the text.
The square of the characteristic strain (directly observable by a long-base
interferometer) il plotted vs. wave frequency ({\it solid line}). {\it
Thick solid line:} {\it LISA} single-arm Michelson sensitivity curve.
{\it Thick dash} at $f\simeq 10^{-9}$: current limits from pulsar
timing experiments. {\it Long-dashed line} at $10^{-6}\lta f\lta 0.1\,$Hz:
expected strain from extragalactic white dwarf binaries (Farmer
\& Phinney 2003). {\it Dotted line} at $10^{-4} \lta f\lta 10^{-3}\,$Hz:
expected strain from unresolved galactic white dwarf binaries (Nelemans
et al. 2001).
}
\label{fig4}
\vspace{+0.5cm}
\end{figurehere}

\section{Results \label{results}}

\subsection{Gravitational Wave Background}

Merger tree realizations allow us to follow the time evolution of individual 
inspiraling binaries, and solve equation (\ref{pinneibis}) by performing a 
direct weighted sum of the emission from individual evolving pairs.
The resulting GWB is shown in Figure 
\ref{fig4}, where the square of the characteristic strain is plotted 
against observed frequency. The peak occurs for $f_{\rm peak}\approx 
10^{-10}$ Hz. For $f\lta f_{\rm peak}$ the spectrum is shaped by the 
evolution of binaries in the stellar slingshot regime, and, according to 
equation (\ref{spectrumSL}), $h_c^2 \propto f^2$ just below $f_{\rm peak}$.
At even lower frequencies the energy distribution is somewhat steeper 
than $f^2$ because of the assumed cutoff in the individual emission spectra 
for $a=a_h$ (see Figure \ref{fig3}). For frequencies $f\gta f_{\rm peak}$, 
but below $10^{-6}$ Hz, $h_c^2 \propto f^{-4/3}$, as expected in the classical 
GW regime (see eqs. \ref{eqhc} and \ref{spectrumGW}). At even higher 
frequencies, $f\gta 10^{-6}$ Hz, the strain is shaped by the convolution 
of the sharp frequency cutoff at the last 
stable orbit of individual binaries (see Figure \ref{fig3}), and in this 
range one has approximatively $h_c^2 \propto f^{-2.6}$. The slope in this 
regime is directly related to the rate of binary coalescences and to the MBH 
mass function. 
\begin{figurehere}
\vspace{0.5cm} \centerline{
\psfig{figure=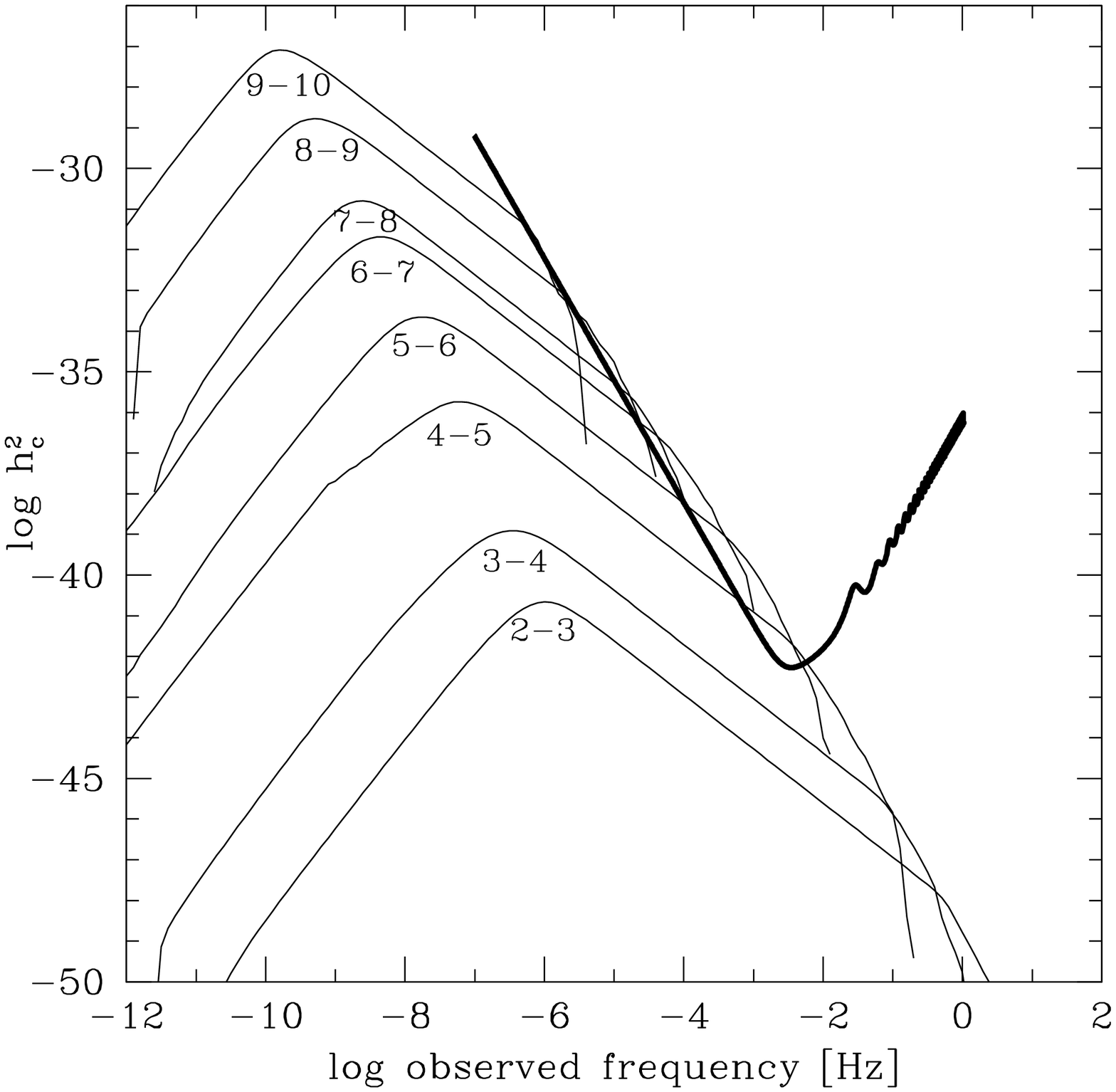,width=3.6in}}
\vspace{-0.0cm}
\caption{\footnotesize Integrated GWB from inspiraling MBH binaries in
different mass ranges. From top to the bottom (as indicated by the labels), the
curves show the signal produced by events with $9<\log(m_1/\msun)<10$\,
$8<\log(m_1/\msun)<9$\,...,$2<\log(m_1/\msun)<3$. Here $m_1$ is the mass of the
heavier black hole.}
\label{fig5}
\vspace{+0.5cm}
\end{figurehere}

Figure \ref{fig4} also shows the {\it LISA} single-arm Michelson sensitivity 
curve for a sky-and-polarization averaged signal with signal-to-noise 
$S/N=1$.\footnote{The 
curve is taken from http://www.srl.caltech.edu/$\sim$shane/sensitivity, where it is 
given in terms of the "effective strain" $h_{\rm eff}=h_c/\sqrt{f}$.} 
It is apparent how {\it LISA} 
could detect the GWB above $10^{-5}\,$Hz, while the peak at nHz 
frequencies is close to current limits of pulsar timing experiments 
(Lommen 2002). We have also 
plotted the GWB expected in the {\it LISA} window from the cosmological 
population of white dwarf binaries computed by Farmer \& Phinney (2003), 
and the GWB from unresolved galactic white dwarf binaries (Nelemans et al. 
2001). Our estimate of the background due to MBH binaries lies well above 
the expected strain from cosmological white dwarf binaries.

The contribution to the GWB from MBH binaries in different mass ranges is 
depicted in Figure \ref{fig5}, while Figure \ref{fig6} shows the signal from 
different redshift intervals. 
From these figures it 
is clear how, for $f\lta 10^{-6}$ Hz, the background is dominated by 
events at moderate redshifts, $0\lta z \lta 2$, 
from very massive binaries, $m_1\gta 10^8\,\msun$. 

\begin{figurehere}
\vspace{-0.5cm} \centerline{
\psfig{figure=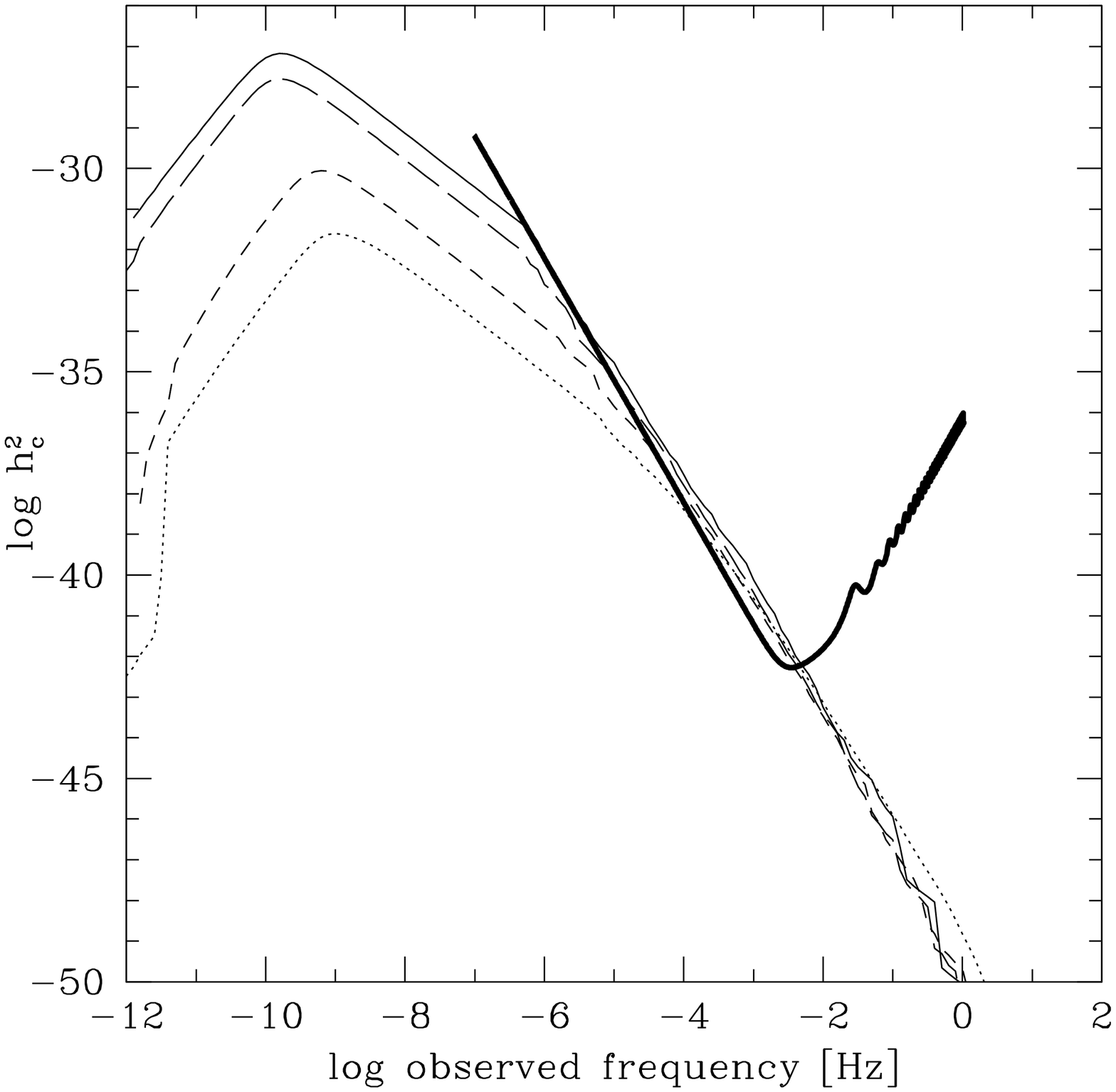,width=3.6in}}
\vspace{-0.0cm}
\caption{\footnotesize GWB from different redshift intervals. From top to
bottom, the spectrum of the characteristic strain is computed integrating the
emission from events at $0<z<1$ ({\it solid line}), $1<z<2$ ({\it long-dashed
line}), $2<z<3$ ({\it short-dashed line}), and $z>3$ ({\it dotted line}), respectively.
}
\label{fig6}
\vspace{+0.5cm}
\end{figurehere}
Lighter binaries show up at higher frequencies, where the background drops in 
amplitude. For $f\gta 10^{-4}$ Hz, the signal is dominated by $m_1\lta
10^{7}\, \msun$ binaries, with comparable contributions from events at $z<3$ 
and $z>3$. The peak frequency $f_{\rm peak}$ of each curve in Figure 
\ref{fig6} is determined by two factors acting in opposite directions, i.e. 
the Hubble expansion and the time evolution of the binary mass function. The 
latter remains approximately constant in the redshift interval $0\lta z \lta2$,
so that the spectrum originating from sources at $1 \lta z \lta 2$ appears, 
because of the Hubble expansion, redshifted with respect to the signal 
from $0 \lta z \lta 1$ events. The peak frequency of events at $z\gta 2$ becomes
bluer because the effect of the reduced average binary mass at increasing 
redshifts dominates over the Hubble expansion.

\subsection{Stochastic versus resolved signal}

Merger tree realizations allow us to compute the GW signal from 
individual MBH binaries, 
and to compare it with the {\it LISA} instrumental noise. Figure \ref{fig7} 
depicts the number of systems above the $S/N=5$ and $S/N=1$ sensivity 
thresholds, per resolution frequency bin ($\Delta f=10^{-8}$ Hz assuming a 
3-year mission duration), and shows that such number is less than unity 
all over the surveyed frequency range. For reference, we also plot the 
total number of sources per bin, i.e. the number of sources regardless of
the strength of their GW signal. Even in this ideal case, for $f\gta 
10^{-3.5}$ Hz, the GWB would be resolved into discrete events.
We have also checked that the signal from the rare resolved events accounts 
for more than the 99\% of the total GWB, i.e. the numerous sources below the 
sensitivity threshold provide a negligible background at Earth. We conclude 
that the GWB from cosmological MBH binaries will be resolved into discrete 
events all over {\it LISA} frequency range. 

Integrating over frequencies, the number of MBH binary systems observable
by {\it LISA} with $S/N$ in excess of 5 (1) is $\approx 20$ (200).  
\footnote{The figures given are obtained neglecting every source of noise other than instrumental. 
In particular, the stochastic noise from galactic WD-WD binaries 
shown in Figure \ref{fig4} (Lommen et al. 2003) has not been considered here.} 
These numbers refer to {\it stationary} events, i.e., binaries at large separation whose GW signal 
does not shift much in frequency, hence longlasting compared to the observation time. The 
number of such events is {\it independent} on any (plusible) mission duration.
On the contrary, the number of observable {\it bursts}, i.e., 
binaries during the observation period, clearly depends on the mission lifetime. 
In $10^8$ secs there will be $\approx 200$ 
bursts (see upper left panel in Figure \ref{fig1}), and among them, 
$\approx 35$ (50) are detectable, by {\it LISA}, with  $S/N>5$ ($S/N>1$).  

\begin{figurehere}
\vspace{-0.5cm} \centerline{
\psfig{figure=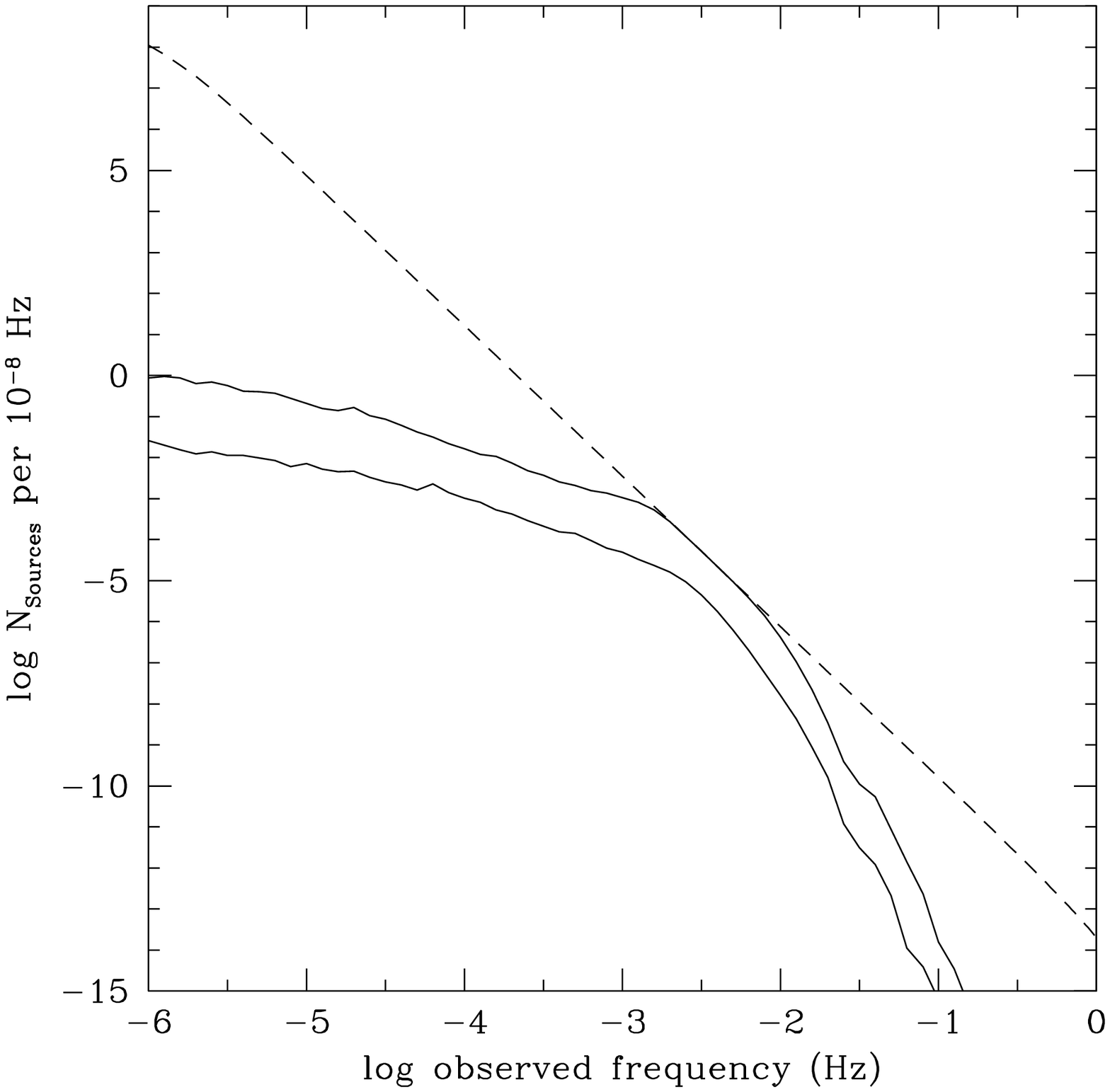,width=3.6in}}
\vspace{-0.0cm}
\caption{\footnotesize Number of MBH binaries per $10^{-8}$ Hz frequency
bin contributing to the GWB as received today. {\it Dashed line:}
all binaries. {\it Upper solid line:} binaries above
{\it LISA} $S/N=1$ sensitivity threshold.
{\it Lower solid line:} binaries above {\it LISA} $S/N=5$
sensitivity threshold.
}
\label{fig7}
\vspace{+0.5cm}
\end{figurehere}

The redshift distribution of detectable events is shown in Figure \ref{fig8}, in the case of 
$S/N=5$ threshold. The distribution of bursts is similar to that of coalescences shown in 
Figure \ref{fig1}, lower left panel, indicating that a relevant fraction of bursts at high 
redshifts produces a signal above threshold. The peak of the redshift distribution of
stationary events is shifted at lower redshift by two different instrumental effects.
First, the inclusion of a detection
threshold selects preferentially binaries at low redshifts. Second, as for wide binaries
the time spent at a given frequency can be larger than the observation time,
only a fraction of the signal (see eqs. 35 and 36)
may actually be detected in a $10^8$ s observation. This fact causes the drop of the
stationary event counts below $z\simeq 3$ seen in Figure \ref{fig8}

\begin{figurehere}
\vspace{-0.5cm}
\centerline{
\psfig{figure=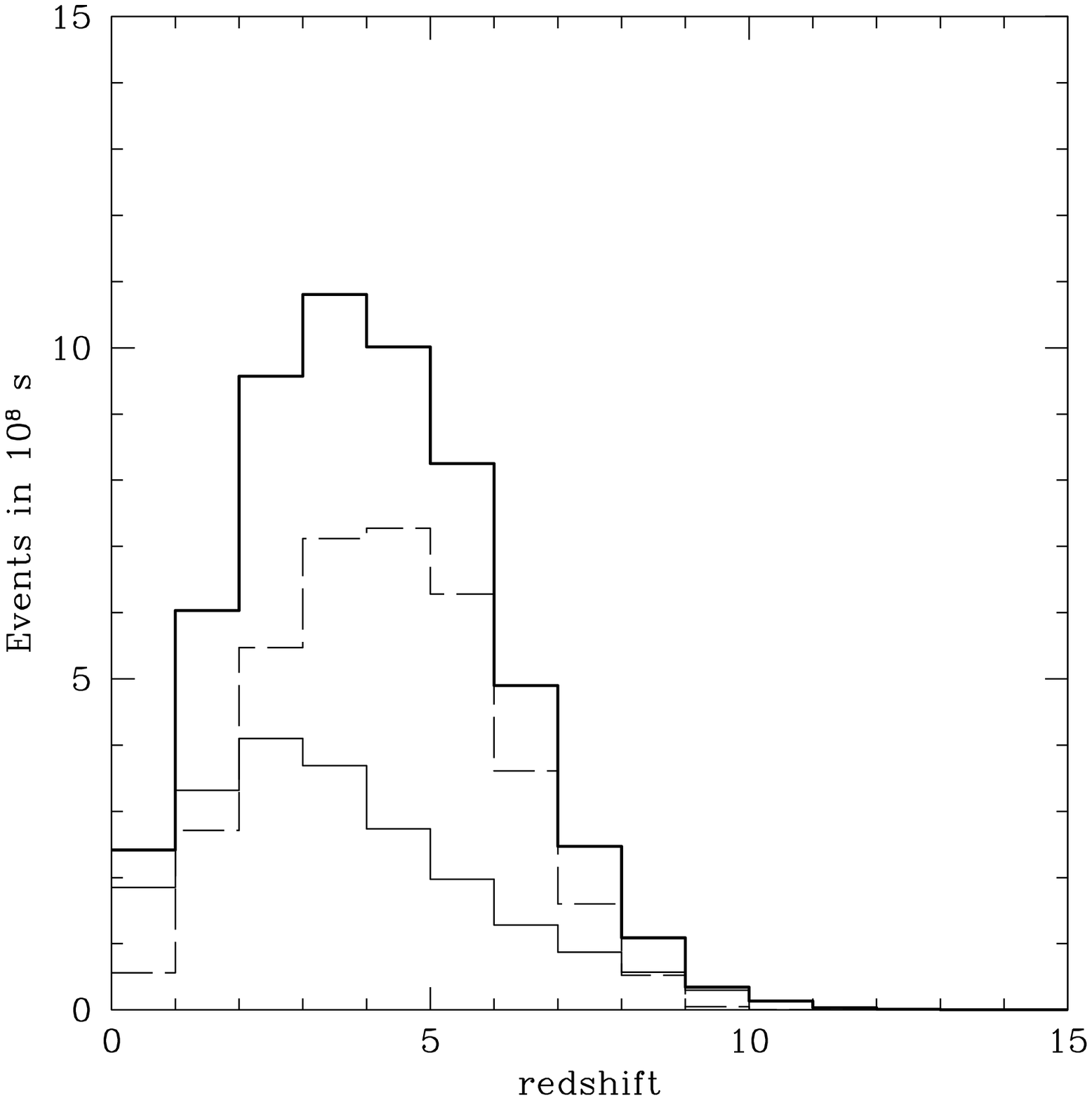,width=3.6in}}
\vspace{-0.0cm}
\caption{\footnotesize
Number of events per unit redshift interval resolved by {\it LISA} with $S/N>5$ in $10^8$ secs. 
{\it Thick-solid histogram:} total number of events in $10^8$ secs. {\it Solid histogram:} number of
stationary events. These events are of much longer duration compared to the mission lifetime.
{\it Dashed histogram}: number of bursts in $10^8$ secs. These events are of short duration
compared to the mission lifetime.
}
\label{fig8}
\vspace{+0.5cm}
\end{figurehere}

\subsection{Test and comparison} 

It is fair at this stage to ask how sensitive our predictions are to 
the details and uncertainties of the adopted scheme for binary formation 
and evolution. To answer this question we have run two test models, arbitrarily
dividing and multiplyng the computed hardening time $t_h$ by a factor of 3, 
i.e. significantly increasing or reducing the rate of binary decay by stellar
dynamical processes. 

In the ``fast'' hardening case, $f_{\rm peak}$ increases by a factor 
$\sim$ 30\% and the peak amplitude drops by $\sim$ 10\%. 
This is due to the fact that, when $t_h$ is shorter, GW emission 
losses dominates over stellar slingshots only at shorter binary separations. 
The opposite occurs in the case of ``slow'' hardening, where $f_{\rm peak}$ is
lowered by a factor $\sim$ 30 \%, and the peak amplitude is $\sim 20\%$ larger than
in the standard case. For $f\gta f_{\rm peak}$, the background amplitude is 
15\% larger in the ``fast'' hardening case, because a larger number of MBH 
binaries can now coalesce. The coalescence rate is now 78 per observed year, to be compared to 
64. Again, the opposite is true in case of ``slow'' 
hardening, the coalescence rate decreasing to 50 per observed year. 
The situation is different in the {\it LISA} window.
A faster or slower hardening has basically no effects, as the increased or 
reduced rate of coalescence affects very massive binaries only, which do not contribute
to the {\it LISA} frequency range. 
Consequently, the precise value of the hardening time has no significant impact on the  
number of sources (either stationary or bursts) observable by {\it LISA}.
This is, of course, only true as long as $t_h$  is less than the then Hubble time. If MBH binaries were
instead to ``stall" (because of the depopulation of the loss cone), the rate of detectable events
could become negligible. We also find our predictions to be sensitive to the occupation fraction of
halos hosting nuclear MBHs. Within our assumptions (rare ``high-$\sigma$" seed holes),
this is of order 50\% at $z\sim 5$ for halos more massive than $10^{10}\, \msun$.
A lower occupation fraction would decrease the rate of inspiraling events, at the
expenses of making MBHs non-ubiquitous in the nuclei of nearby galaxies. 
The black hole coalescing rate would be higher if the seeds were more common at early times.

We have compared our results with the calculations of Wyithe \& Loeb
(2003), who estimated the GWB from MBH binaries using a different, 
but similar in spirit, approach. In their work the time dependent binary mass function 
is explicitly derived from the extended Press-Schechter formalism. The nature 
of such methodology hampers the possibility of following the evolution of 
individual MBH pairs (e.g., triple interactions as well as the differential 
frequency broadening due to the non-negligible duration of the orbital decay 
phase are neglected). The two estimates of the GWB
appear in good agreement (to within a factor 2) at high frequencies
($f\gta 10^{-4}$ Hz), a fact reflecting the
common hierarchical scenario, the similar normalization to the present-day
$m_{\rm BH}-\sigma_*$ relation, and our relatively short binary decay
timescales. By contrast, our predicted number counts are about an order of magnitude
smaller than given by Wyithe \& Loeb (2003). This is because they do not
consider the shift in frequency during the inspiraling phase, and compute
the counts at a fixed frequency of $10^{-3}$ Hz. Their simplified treatment has the effect
of artificially boosting (by as much as a factor of $\sim 50$)
the number of detectable events at $z \gta 7$.

Lastly, we were concerned that, in the {\it LISA} 
window, the GWB we computed is entirely due to GWs emitted at the last stable orbit
of the coalescing pair, a regime the Newtonian approximation may fail to 
describe adequately. To quantify this we have implemented a post-Newtonian
(order 2) approximation describing the GW emission from binary
systems (Blanchet 2001). We find that differences in the
spectrum of a coalescing MBH binary are quite small, of order 10\%, and the 
overall results are, basically, unaffected.

\section{Summary and conclusions}

We have computed the gravitational wave signal (in terms of the characteristic 
strain spectrum) from the cosmological population of inspiraling MBH binaries 
predicted to form 
at the center of galaxies in a hierarchical structure formation scenario.
The assembly and growth of MBHs was followed using Monte Carlo realizations of the 
halo merger hierarchy from the present epoch to very high redshifts, coupled 
with semi-analytical recipes treating the dynamics of the (inevitably 
forming) MBH binaries. We find that the broad band GWB spectrum can be 
divided into 
three different regimes. For frequencies $\lta 10^{-10}$ Hz, the GWB is shaped by 
MBH binaries in the gravitational slingshot regime, i.e. the orbital decay is driven 
by scattering off background stars. In the intermediate band, $10^{-9}\lta f 
\lta 10^{-6}$ Hz, GW emission itself dominates potential energy losses and 
the strain has the ``standard'' $f^{-2/3}$ behaviour. Finally, for $f\gta 10^{-6}$, 
the background signal is formed by the convolution of the emission at the last 
stable orbit from individual binaries. In the first two regimes, the strain is 
dominated by low redshift events ($z\lta 2$) involving genuinely supermassive holes. 
At larger frequencies the main contribution comes from lighter and lighter 
binaries. Such background should dominate that predicted from a 
population of extragalactic white dwarf binaries. As already pointed out by Wyithe 
\& Loeb (2003), in the nHz regime probed by pulsar timing experiments, the amplitude 
of the characteristic strain from coalescing MBH binaries is close to current 
experimental limits (Lommen 2002).

In the {\it LISA} window ($10^{-6}\lta f\lta 0.1$ Hz), 
the main sources of GWs are MBH binaries in the mass range $10^{3}\lta m_1\lta 
10^{7}$ $\msun$. With a plausible lifetime of $\simeq 3$ years, {\it LISA} will resolve the GWB 
into $\approx 60$ discrete sources above $S/N=5$ confidence level. 
Among these, $\approx 35$ are bursts, i.e., 
binaries caught in their final inspiraling phase. We predict that most of the 
observable events will be
at redshifts $2 \lta z \lta 7$, with only a handful at larger redshifts.
While {\it LISA} will make it possible to probe the coalescence of
early black hole binaries in the universe, it may not be able to observe
the formation epochs of the first MBHs.

\acknowledgments
\ni We have benefitted from several discussions with V. Gorini. Support for 
this work was provided by NASA grant NAG5-11513 and NSF grant AST-0205738 
(P.M.).

{}

\end{document}